\documentclass[12pt]{cernprep}
\usepackage{graphicx}
\usepackage{amsmath}
\usepackage{amssymb}
\usepackage{epsfig,color,rotating}
\usepackage{epstopdf}
\usepackage{thumbpdf}
\usepackage[pdftex,
        pdftitle={A Document With An Image},
        pdfauthor={Udo K. Schuermann, Ringlord Technologies},
        pdfsubject={A Demonstration with Extra Cheese},
        pdfkeywords={latex pdf gimp batch thumbnails hyperlinks},
        pagebackref,
        pdfpagemode=None,
        bookmarksopen=true]{hyperref}
\usepackage{euscript}
\usepackage{multirow}
\usepackage{colortbl}
\usepackage{calc}

\makeatletter
\newlength\twolinebox@linelength
\newlength\twolinebox@columnheight
\newcommand{\twolinebox}[2]{%
   \setlength{\twolinebox@linelength}%
             {\maxof{\widthof{#1}}{\widthof{#2}}}%
   \setlength{\twolinebox@columnheight}{\heightof{#1}+\depthof{#1}+0.2em+0.4em/2+\heightof{0}/2}%
   \raisebox{0pt}[\twolinebox@columnheight][\heightof{\vbox{\vskip0.2em\hbox to 
   \twolinebox@linelength {#1\hfil}\vskip0.4em\hbox to 
   \twolinebox@linelength {#2\hfil}}}+\depthof{\vbox{\vskip0.2em\hbox to 
   \twolinebox@linelength {#1\hfil}\vskip0.4em\hbox to 
   \twolinebox@linelength {#2\hfil}}}-\twolinebox@columnheight+0.2em]{\vbox to 
   \twolinebox@columnheight{\vskip0.2em\hbox to 
   \twolinebox@linelength {#1\hfil}\vskip0.4em\hbox to 
   \twolinebox@linelength {#2\hfil}}}%
}
\newcommand{\tw}      {\textwidth}

\newcommand{\qbar}    {{\bar q}}

\def\slashii#1{\setbox0=\hbox{$#1$}            
  \dimen0=\wd0                                 
  \setbox1=\hbox{\sl/} \dimen1=\wd1            
  \ifdim\dimen0>\dimen1                        
     \rlap{\hbox to \dimen0{\hfil\sl/\hfil}}   
     #1                                        
  \else                                        
     \rlap{\hbox to \dimen1{\hfil$#1$\hfil}}   
     \hbox{\sl/}                               
  \fi}

\definecolor{rltbrightred}{rgb}{1,0,0}
\definecolor{rltred}{rgb}{0.75,0,0}
\definecolor{rltdarkred}{rgb}{0.5,0,0}
\definecolor{rltbrightgreen}{rgb}{0,0.75,0}
\definecolor{rltgreen}{rgb}{0,0.5,0}
\definecolor{rltdarkgreen}{rgb}{0,0,0.25}
\definecolor{rltbrightblue}{rgb}{0,0,1}
\definecolor{rltblue}{rgb}{0,0,0.75}
\definecolor{rltdarkblue}{rgb}{0,0,0.5}
\definecolor{webred}{rgb}{0.5,.25,0}
\definecolor{webblue}{rgb}{0,0,0.75}
\definecolor{webgreen}{rgb}{0,0.5,0}
\definecolor{Black}{rgb}{0,0,0}
\definecolor{Greymax}{rgb}{0.65,0.65,0.65}
\definecolor{Greycen}{rgb}{0.75,0.75,0.75}
\definecolor{Greymin}{rgb}{0.85,0.85,0.85}
\definecolor{hl}{rgb}{
                0.909803922,       
                0.82745098,               
                0.909803922}
\begin{document}

\begin{titlepage}
 
\vspace{15mm}
\begin{center}
{\LARGE\bf 
Ascertaining the origin of the \ $l\nu l\nu$ excess events \\ 
\vspace{6mm}
at the LHC  by a change of beam energy
} \\      
\end{center}

\vspace{25mm}


\begin{center}
{\large\bf  Mieczyslaw Witold Krasny } \\

\vspace{5mm}

LPNHE, Universit\'{e}s Paris VI et VII and CNRS--IN2P3, Paris, France \\


\end{center}

\vspace{30mm}
\begin{abstract}
A higher than predicted rate of two leptons plus missing transverse energy events, 
reported at the summer HEP conferences,  can originate from 
a decay of the Higgs boson into a $WW^{(*)}$ pair, 
a misjudgement of the rate of SM background processes or 
a  statistical fluctuation. 
In this paper  we discuss a way to resolve this three-fold ambiguity.  

\end{abstract}

\vspace{15mm}

\end{titlepage}

\section{Introduction}

The LHC and the Tevatron experiments are trying to close  the window for the Standard Model (SM)
 Higgs in the low mass region,  favored by the precision measurements of the SM parameters.

The question discussed here is if, and how, the LHC experiments can firmly establish the existence, or exclude,
by the end of 2012,  the SM Higgs boson in the mass region  of 120--150 GeV/c$^2$. 
This mass region is of particular interest because the excess events reported at the Grenoble EPS Conference\cite{Murray} 
can be interpreted as coming from the $ H \rightarrow WW^{(*)}$ decays. Both the ATLAS and CMS collaborations  
reported an excess of events with respect to the SM expectation  in the above mass range. 
 %
%

At the Lepton-Photon Conference in Mumbai,  the CMS collaboration\cite{Sharma} and the ATLAS collaboration\cite{Nisati}
presented updated results corresponding to 1.5 fb$^{-1}$ and  1.7 fb$^{-1}$ of integrated luminosity, respectively. 
The statistical significance of the excess was reported to be reduced  to the  ~2$\sigma$ level (ATLAS)  
and to  the  ~1$\sigma$ level (CMS).

The LHC machine has reached its stable operation mode with the 50 ns bunch spacing 
delivering of the order of 40 pb$^{-1}$ per day. 
If such a performance is maintained during  the $pp$ running periods in 2011 and 2012
each experiment may collect of the order of 15 fb$^{-1}$ of integrated luminosity.

If the present excess of events represents a statistical fluctuation, 
the expected  increase of the statistical precision in 2011 and 2012 will be largely 
sufficient to reject firmly the existence of Higgs boson in the discussed mass range. 
On the other hand, if the excess of events is confirmed with high statistics data,  the following two hypothesis
will remain to be resolved:  \\
\begin{enumerate}
\item 
the excess events originate from the Higgs boson decays. \\
\item 
the excess events 
reflect higher than expected rate of SM background processes, \\
\end{enumerate}

This will be difficult because the statistical errors will no longer be dominating. 
If the observed, $N_{obs}$,  and the expected background and signal, $N_{bgr}$, $N_{Higgs}$,  numbers  of events reported by the ATLAS collaboration at the Lepton-Photon Conference for the 0 jet selection, and for $M_{Higgs} = 150$ GeV/c$^2$ are scaled up 
to the integrated luminosity of 15 fb$^{-1}$ and the two following assumptions are made: 
\begin{itemize} 
\item
the relative errors on the numbers 
of the expected  SM  background and Higgs  events will not be improved, 
\item
the present excess number of observed events is genuine (representing its infinite luminosity asymptotic value), 
\end{itemize} 
then $N_{obs} = 618 \pm 24$,  $N_{bgr} = 465 \pm 78$ and  $N_{Higgs} = 300 \pm 60$. 
It is thus obvious that the  hypothesis 2 can be confirmed only at  the 2$\sigma$ level and the hypothesis 1 can be tested only 
at the 1.5$\sigma$ level. 
The errors on $N_{bgr}$ and $N_{Higgs}$ will have to be reduced  by at least a factor of 3
(i.e. they will have to evolve  with the collected  luminosity $L$ as ~$1/\sqrt{L}$) in order 
to firmly reject or to establish the existence of the SM Higgs boson. 
This will be anything but simple because the errors
are  dominated  entirely by the theory and Monte-Carlo modeling uncertainties\cite{Cranmer}
which give rise to irreducible errors. They may be reduced  with increasing luminosity  but certainly less than the experimental measurement errors, for which  the ~$1/\sqrt{L}$ evolution reflects  already a rather  optimistic 
scenario.  

In this note we shall discuss the measurement strategy capable to  bypass the 
dominant systematic modeling uncertainties. We shall  exploit the difference
in the production mechanism of  the background and the Higgs events and propose an observable 
capable to firmly identify the source of the excess events, provided that  the data are taken 
at the LHC at two different beam energies. 
The  aim  of the strategy presented here is to ascertain the origin of the excess events (if it remains) independently 
of the progress in reducing the modeling and theoretical uncertainties.

\section{The observable} 

\begin{figure}[h]
\begin{center}
\includegraphics[width=0.90\textwidth]{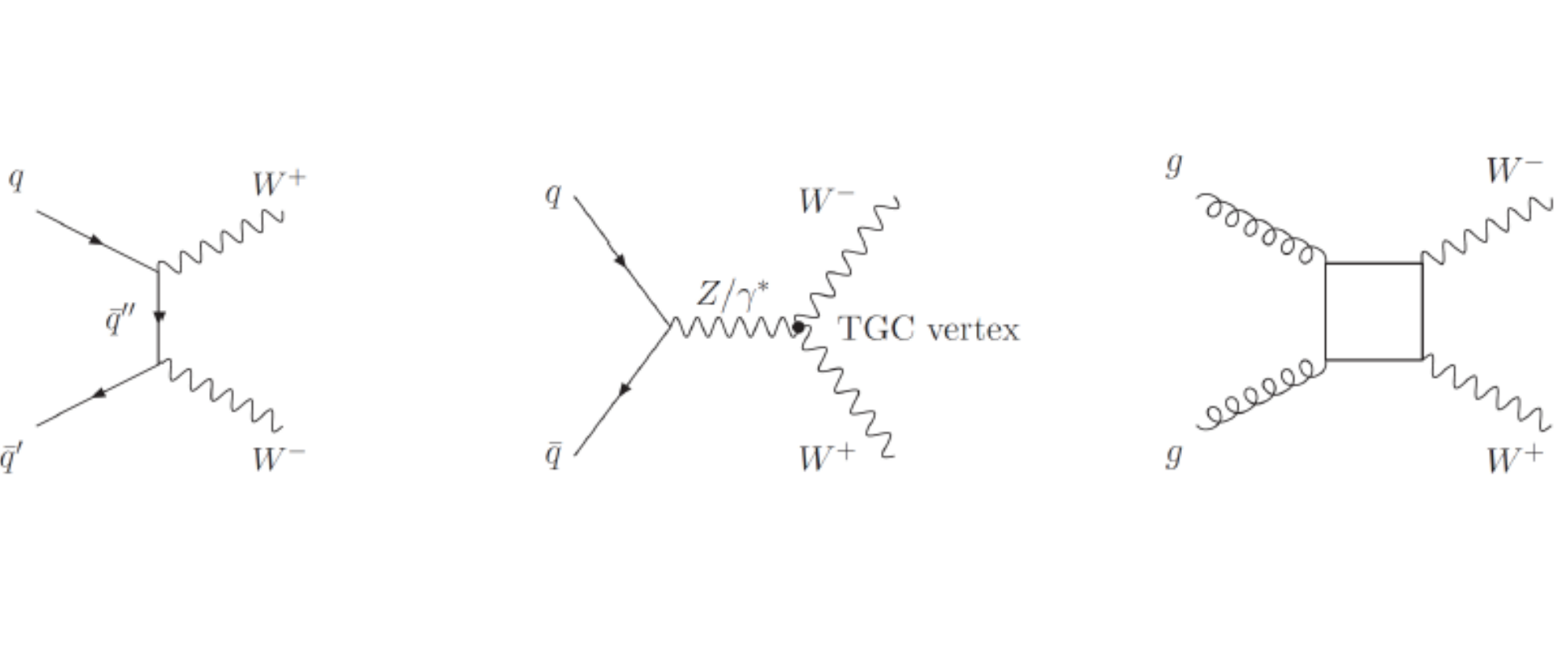} 
\end{center}
\caption{The dominant non-resonant $WW$ pair production diagrams. In 
the mass region studied in this paper one of the $W$-bosons is virtual.}
\label{WWcontinuum}
\end{figure}

Let us focus our attention on the 0 jet subsample of the $l\nu l\nu$ events\cite{Murray}. 
For this subsample, the source of background events  for the Higgs searches is predominantly a non-resonant 
production of the $WW$ pairs,  depicted in Fig.~\ref{WWcontinuum}. 
The processes which dominate are the quark-antiquark collisions.  
Collisions of  gluons contribute to the event rate at the level of  $\sim$ $3 \%$.

\begin{figure}[h]
\begin{center}
\includegraphics[width=0.90\textwidth]{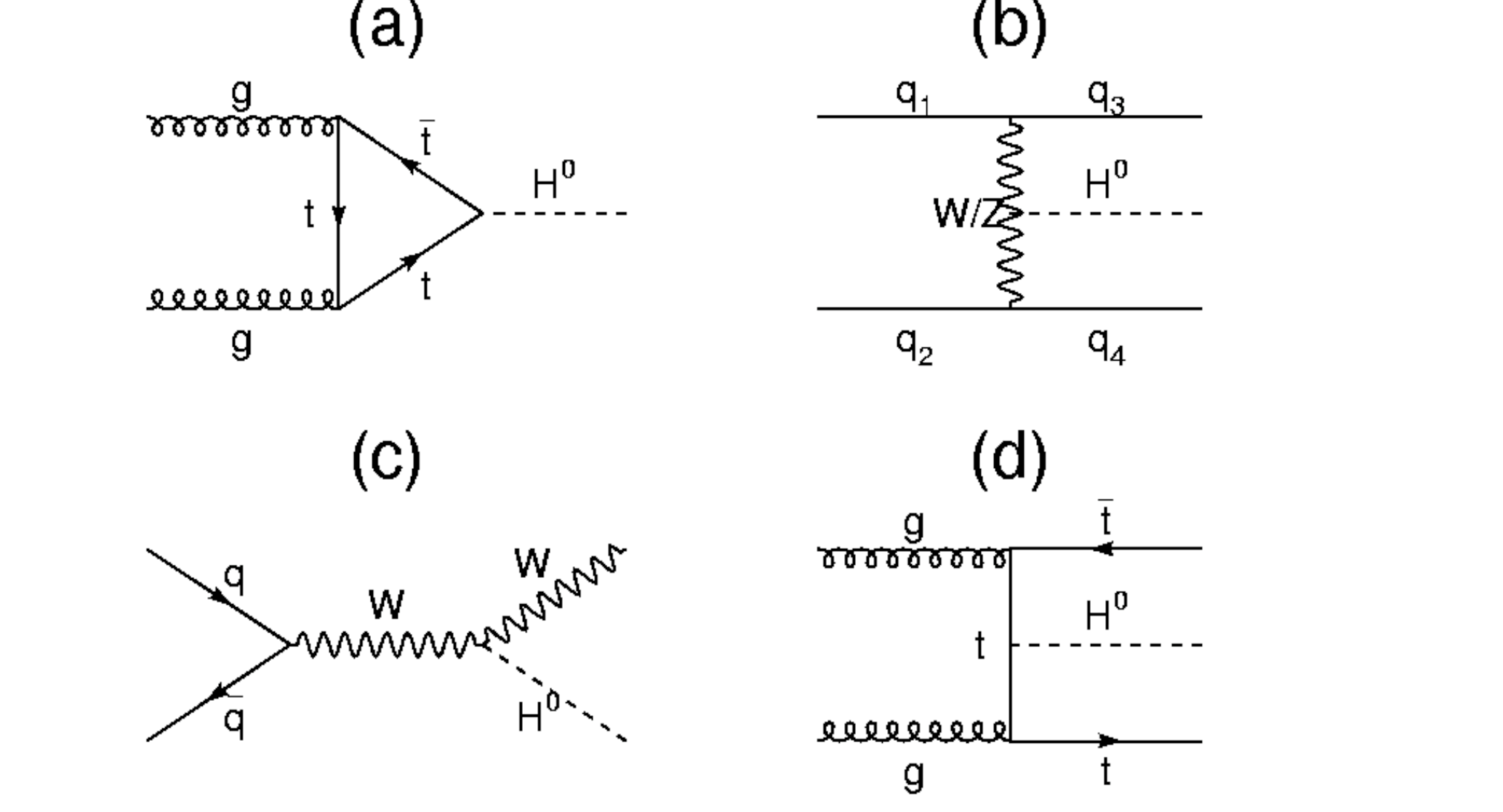} 
\end{center}
\caption{The dominant Higgs production diagrams.}
\label{WWresonant}
\end{figure}

If the SM Higgs boson exists, the  $WW^{(*)}$ pairs are also coming from the $H \rightarrow WW^{(*)}$ decays. 
In the discussed mass range, the Higgs boson is produced predominantly  
in gluon-gluon collisions. 
The contribution of the quark-initiated processes (b) and (c), 
depicted in Fig.~\ref{WWresonant}, 
is at the level of $\sim$ $10 \%$. 

The relative magnitude of the Higgs and of the SM background contributions to the observed even rates
could thus be established by measuring the relative strength of the  gluon-gluon collision processes 
with respect to the quark-antiquark ones. Of course, these processes cannot be distinguished 
on  the event-by-event basis.  However, their relative strength can be changed  
by modifying the centre-of-mass energy of colliding beams. This is the main idea presented in 
this note.

For simplicity of arguments let us consider the central (zero-rapidity) 
production of the $WW^{(*)}$  pairs with the  invariant mass  $m_0 = 150$  GeV/c$^2$, 
in the simplified framework  based on collinear, massless partons. 
If the centre-of-mass-energy-squared of $pp$ collisions changes from $s_0$ to $s_1$,  the momentum fraction 
of partons producing exclusively the $WW^{(*)}$ pairs  changes from $x_0=\sqrt{m^2_0/s_0}$ to $x_1=\sqrt{m^2_0/s_1}$. 

If protons were composed only of gluons and sea quarks,  the relative magnitude of the gluon- and quark-initiated
processes could not be resolved by measuring the rates   at the two $s$ values,  because 
the ratio of the sea quark fluxes  at $x_0$ and $x_1$ is to a good approximation the same as the ratio 
of the gluon fluxes. This is a direct consequence of the DGLAP evolution equation.    
In the discussed range of  $m^2_0 \gg \Lambda _{QCD}$ and in the LHC  range of $s$ 
the non-perturbative differences in the $x$-shape of the sea quark and gluon distributions are 
washed out when evolved to the   $Q^2 = m^2_0$ scale. 

The quark and gluon initiated processes can be resolved because of the presence of the valence quarks 
in the protons.   

\begin{figure}[h] 
  \begin{center}
    \includegraphics[width=0.495\tw]{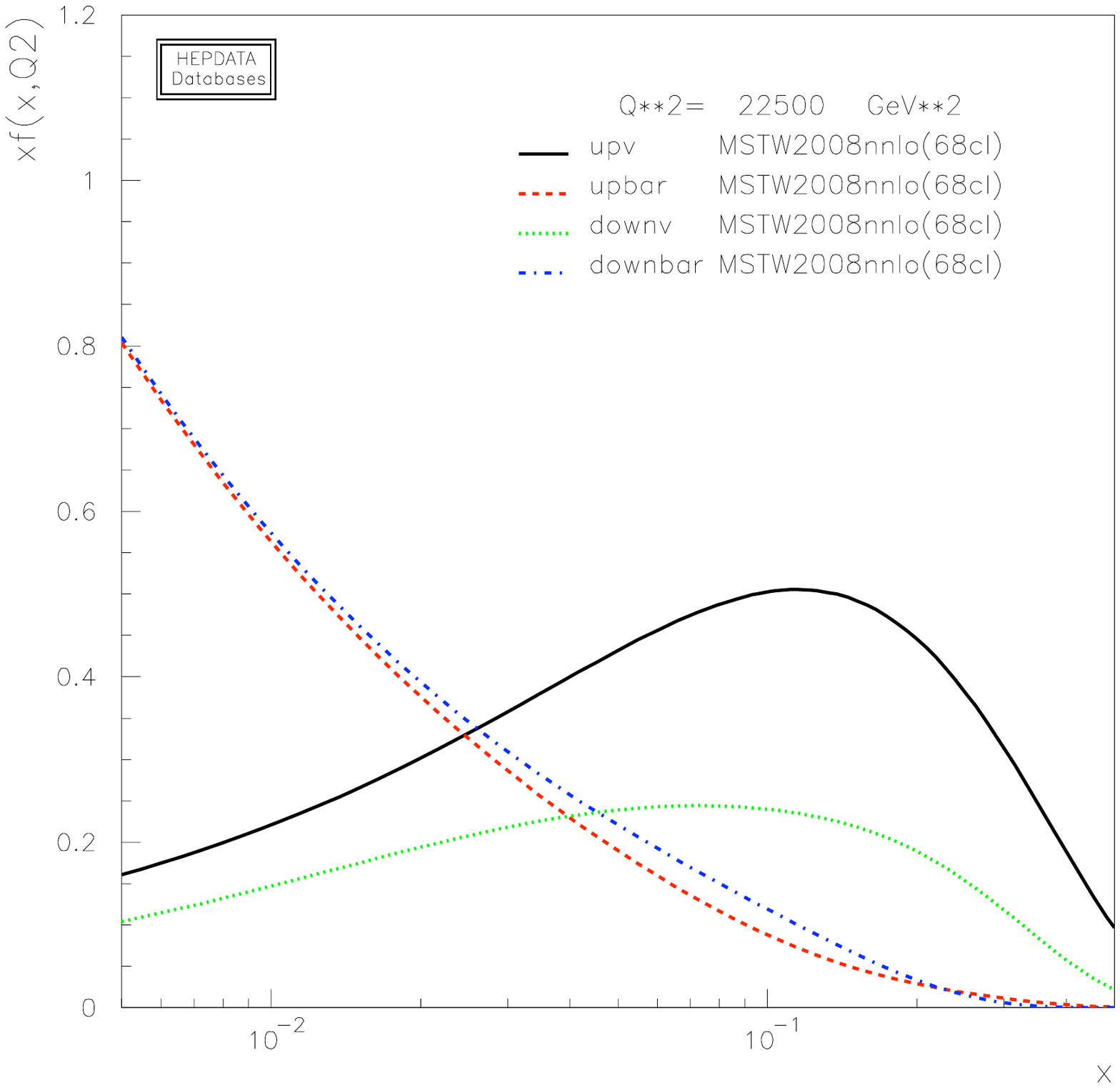}
    \hfill
    \includegraphics[width=0.495\tw]{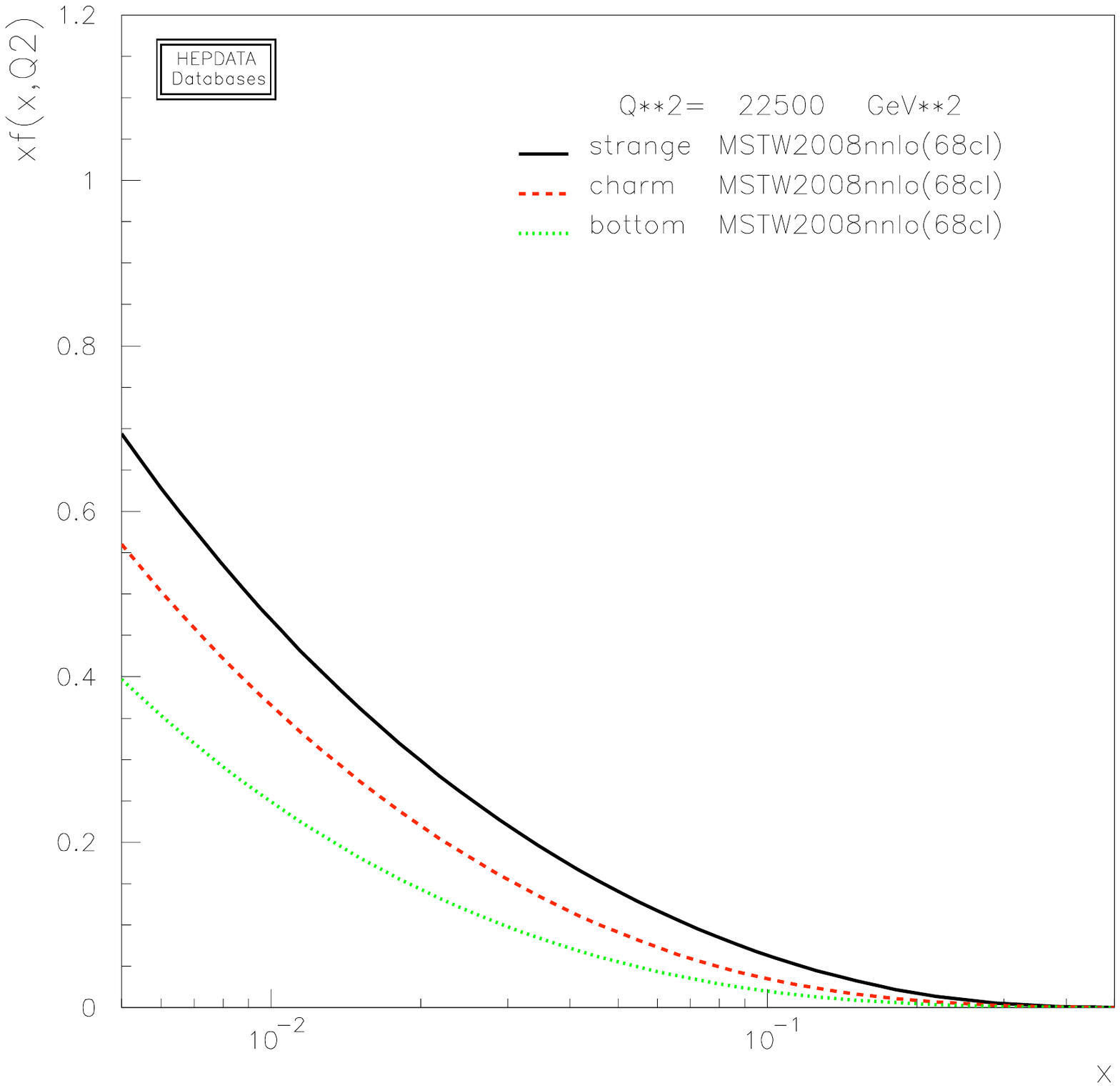}
    \caption{The parton distribution functions of the valence and sea quarks at $Q^2 = 22500$ GeV$^2$.}               
    \label{quarks}
  \end{center}
\end{figure}

For  $x \sim 10^{-2}$, corresponding at the LHC energies to the discussed mass region,  
the  valence quark fluxes decrease with decreasing $x$, contrary to the sea-quark and  gluon fluxes which 
strongly increase with decreasing $x$ value.  This is illustrated in Fig.~\ref{quarks},  where the distributions 
of the valence and sea quarks are  shown as a  function of $x$ at the  $Q^2 = 22500$ GeV$^2$ scale\footnote{This and 
all the subsequent  plots showing partonic densities were made using  the Durham HepData Project Tool \cite{Durham}.}. 
The difference in the $x$-dependence of the 
quark and gluon/sea quark fluxes  allows to change  the relative proportion of the gluon and the quark initiated processes
by modifying  the energy of the LHC beams\footnote{It is a very lucky coincidence that, for the LHC beam 
energies,  the resolving power happens to be maximal in the mass region where the excess of events is seen.}. 

Motivated by the above considerations,  
we propose to use the following observable to measure  the relative contribution  of the gluon- and of the quark-initiated processes:
\begin{eqnarray}
R(s_0,s_1, m_0) = \frac {\sigma (s_1, m_0, E^T_{jet})}{ \sigma (s_0, m_0, E^T_{jet} )},
\end{eqnarray}
where 
 $\sigma (s_0, m_0, E^T_{jet})$ and  $\sigma (s_1, m_0, E^T_{jet})$ are  the integrals of the differential cross sections \\
 $d \sigma (m_t, s_0, m_0, E^T_{jet}) / dm_t$  and  $d \sigma (m_t, s_1, m_0, E^T_{jet}) / dm_t$ integrated over 
the region $(0.75 \times m_0 < m_t < m_0)$
of the transverse mass, $m_t$, of the two charged lepton and two neutrino system,   and $E^T_{jet}$ is the  jet energy cut-off 
used in the selection of the 0-jet subsample events.  
For the quark-initiated processes, the  central production of the $WW^{(*)}$ pair ($y_{WW}=0$), $E^T_{jet} \ll m_0$, and in the absence of the higher-twists effects,   $R(s_0,s_1,m_0)$ can be written in the Born approximation as:
\begin{eqnarray}
R(s_0,s_1, m_0) \sim  R^{q \qbar}(s_0,s_1, m_0) =
 \frac {\Sigma _f q_f (\sqrt{m_0^2/s_1},m_0^2) \qbar _f (\sqrt{m_0^2/s_1},m_0^2 )} {  {\Sigma _f q(\sqrt{m_0^2/s_0}, m_0^2 ) \qbar _f (\sqrt{m_0^2/s_0}, m_0^2)}}.
\end{eqnarray} 
For the gluon initiated processes  $R(s_0,s_1,m_0)$ can be written in the Born approximation as: 
\begin{eqnarray}
R(s_0,s_1, m_0) \sim  R^{gg}(s_0,s_1,m_0)  = \frac {g^2 (\sqrt{m_0^2/s_1},m_0^2 )} {  { g^2 (\sqrt{m_0^2/s_0},m_0^2)}}.
\end{eqnarray} 
In the above formulae $q_f(x, Q^2)$, $\qbar _f(x, Q^2)$ and $g(x, Q^2)$ denote the flavour $f$-dependent  quark, antiquark
and gluon distribution functions (PDFs).

The merit of measuring $R(s_0,s_1,m_0)$  is twofold.

\begin{figure}[h] 
  \begin{center}
    \includegraphics[width=0.495\tw]{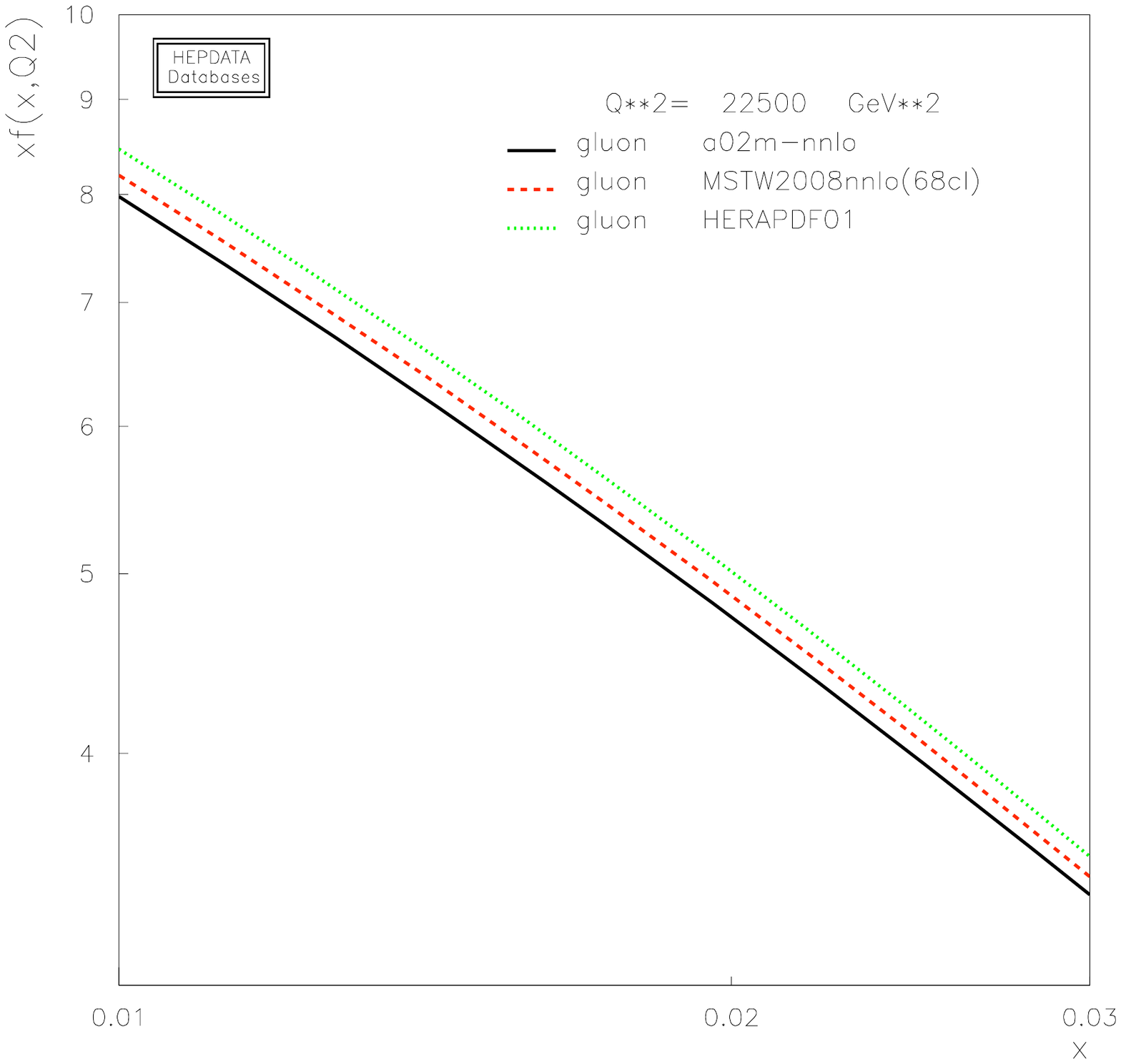}
    \hfill
    \includegraphics[width=0.495\tw]{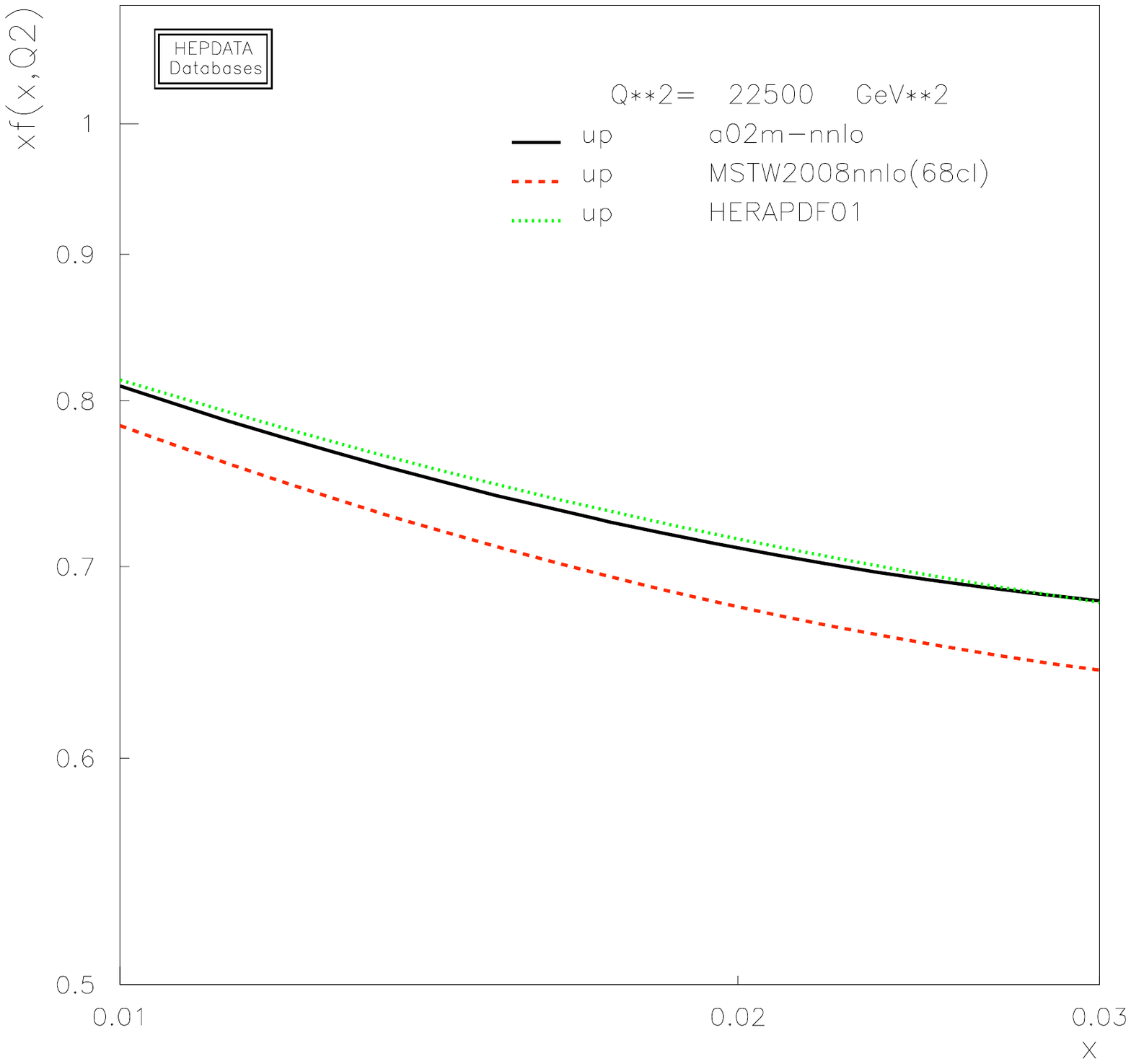}
    \caption{The gluon and the up-quark PDF sets. Note, that the PDF ratios taken at two $x$ values are, to a large extent, 
     independent of  PDF set.}               
    \label{gluons}
  \end{center}
\end{figure}

\begin{enumerate}
\item
From the experimental point of view,  majority  systematic measurement uncertainties
 cancel in the ratio for a stable detector,   if runs taken at two different energies have similar distribution of the number of collisions per bunch-crossing, 
 and if the $s$-dependence of the effects due to co-moving partons are experimentally controlled.  Such an observable is, in particular,  less sensitive to the absolute scale of the lepton
and jet energies. The dominant systematic error for the measured ratio reflects the uncertainty of the relative 
normalization of the data samples taken at the two centre-of-mass energies. For the present method, based on the van der Meer scan\cite{ATLAS_LUMI}, the expected error on the ratio will be of the order of $5 \%$. 
This error can be diminished  by a factor of about $3$ by making use of the well-known $Z$ 
production cross-section ratio, $\sigma^Z_{th}(s_1)/\sigma^Z_{th}(s_0)$,  and measuring, instead 
of $R(s_0,s_1,m_0)$,  the ratio   $R_Z(s_0,s_1,m_0)$ defined as:
\begin{eqnarray}
R_Z(s_0,s_1,m_0) = \frac {N (s_1, m_0, E^T_{jet})}{ N(s_0, m_0, E^T_{jet} )} \times 
\frac {dN/dm_{ll} (s_0,m_{ll}=M_Z)}{ dN/dm_{ll} (s_1,m_{ll}=M_Z )} \times 
\frac {\sigma^Z_{th}(s_1)}{\sigma^Z_{th}(s_0)}, 
\end{eqnarray}
where $m_{ll}$ is the invariant mass of the opposite charge,  same flavour lepton pairs and  $M_Z$ is 
the $Z$-boson mass\footnote{In the future,  the relative luminosity error  could be reduced to a per-mille level if 
luminosity measurement method proposed in \cite{krasny_lumi} is implemented.}.   
\item
The principal merit of the proposed  ratio  is, however, its robustness 
 with respect to the theoretical/phenomenological modeling uncertainties.
$R(s_0,s_1,m_0)$, can be directly interpreted in terms of: $R^{q \qbar}(s_0,s_1,m_0)$, 
$R^{gg}(s_0,s_1,m_0)$ and  the ratio of the valence-quark  to  sea-quark  PDFs.  
We have  found  that the sensitivity to the assumed form of the PDFs  (NNPDF, ABKM. MSTW, CTEQ)\cite{Durham},  is reduced by 
at least a factor of 10 for $R^{q \qbar}(s_0,s_1,m_0)$  and  $R^{gg}(s_0,s_1,m_0)$ with respect to the 
fixed $s$ analysis. This may be easily understood by looking at  
Fig.~\ref{gluons},  where the gluon and up-quark distributions are shown in the relevant $x$-range\footnote{For $m_0 = 150$ GeV and the LHC beam energies of 2.5, 3.5  and 
4.5 TeV the corresponding $x$ values are $x= 0.03,  0.021, 0.017$.}, correspondingly.
While the PDFs differ in normalization,  their ratios taken at the $x_0$ and $x_1$ values 
are  independent of the PDF set.   
The sensitivity to the PDFs  errors is thus restricted solely to our present understanding of the ratio of the valence to the sea quarks, 
which is known presently within a  $\sim$ $5 \%$ uncertainty\footnote{An experiment has been proposed at the CERN SPS to 
improve the precision of this ratio \cite{Dydak}.}. 
It remains  to be added that the proposed observable becomes largely insensitive to the missing higher-order QCD corrections 
(more precisely, to those of them that  are similar for the gluon-gluon and quark-quark initiated processes). 
\end{enumerate} 
 
\section{LHC running scenarios and their resolving power} 

The minimal requirement  which allows to measure the relative contribution of the 
gluon-gluon and quark-antiquark collision processes to the observed event rates is 
to collect the data  at  two different beam energies. We have evaluated numerically two running scenarios, each of them 
for the integrated luminosity of 15 fb$^{-1}$. In the first one 6  fb$^{-1}$ is collected with the 3.5 TeV proton beams and 
9  fb$^{-1}$ at 2.5 TeV. The second one corresponds to 9  fb$^{-1}$ collected with the 3.5 TeV  beams 
and 6  fb$^{-1}$ at 4.5 TeV. 

The relative luminosity in these scenarios minimize the statistical uncertainty 
on the  \\ $R(s_0,s_1,m_0)$ ratio. In the following,  $s_0$ corresponds to the present beam energy and $s_1$
to the reduced (increased) energy for the scenario 1 (2). 

In the estimations presented in this section we have used the signal and background rates, 
following all the experimental cuts,   in the 0-jet channel 
presented by the ATLAS collaboration  at the Lepton-Photon conference \cite{Nisati}. 
The evaluation was made for the  Higgs boson mass of 150 GeV/c$^2$. 
We have assumed further that all the $WW^{(*)}$ pairs are produced centrally and exclusively. 
We have neglected the small $gg$ contribution to the $WW^{(*)}$ background and the small $q \qbar$ contribution to the Higgs 
production process.  These approximations can be abandoned in a technically more advanced analysis,
by including    the realistic detector acceptance for the the $l\nu l\nu$ events and by getting rid of approximations 
made in the presented calculations. This would be obligatory for the realistic analysis of the data
but not necessary in the evaluation of the resolving  power of the method presented in this note.   

For  the first running scenario  $R^{gg}(s_0,s_1,m_0) = 0.56 \pm 0.02$,  and 
$ R^{q \qbar}(s_0,s_1, m_0)= 0.71 \pm 0.02$.  
The measurement of $R(s_0,s_1,m_0)$ by the ATLAS and CMS experiments would thus provide 
a model-independent, discrimination at the 2$\sigma$ level\footnote{For the running scenarios presented above the dominant source of uncertainty  on $R(s_0,s_1,m_0)$ is of statistical 
nature -- if the current  event selection procedures are  maintained for the analysis of the full data sample. 
There is a room for an improvement here by using  less restrictive experimental cuts. The optimal procedure 
would be to analyze  $R(s_0,s_1,m_0)$ in terms of the relative yield of the gluon and quark originated
processes at each stage of the event selection chain corresponding to  their variable mixture. Using 
such a procedure the statistical errors will be reduced and a better understanding of the resolving power of the proposed
method will  be achieved. It has to be stressed that for such a procedure  the relative luminosity uncertainty  measured 
with the van der Meer method will become  a dominant one. A remedy proposed in this paper, 
adequate for the discussed luminosity range  is to replace the measurement of $R(s_0,s_1,m_0)$ by a measurement of $R_Z (s_0,s_1,m_0)$.} between the hypothesis 1 (Higgs + SM background)
and the hypothesis 2 (SM background only). 
This additional (with respect to the current method based on the absolute rates of events)   
discrimination power would be  decisive  to confirm or reject firmly the Higgs boson hypothesis 
if the current  uncertainties of  the expected {\it absolute} signal and background rates are not reduced by a factor of $3$.

For the second running scenario 
 $R^{gg}(s_0,s_1,m_0) = 1.51 \pm 0.03$ 
and $R^{q \qbar}(s_0,s_1, m_0) = 1.28 \pm 0.03$. 
The resolving power of the gluon against the  quark-initiated processes is slightly reduced
due to a smaller contribution of the valence quarks to the overall $q \qbar$ fluxes. 
The discriminating  power of the $R(s_0,s_1,m_0)$ measurement between the 
hypothesis 1 and  the hypothesis 2 stays, however,  at the same level 
because  the reduction of the resolving power 
of the gluon- and quark-initiated processes is compensated by the gain  in the total number 
of both the signal and background events. 

\section{Outlook}

The arguments presented in this note would be irrelevant, while considering the running scenarios
in 2012, if the excess of events disappeared by the end of this year. If it persists,  ascertaining 
experimentally the origin
of the excess events,  no matter what progress will be made in improving the precision of calculation of 
the signal and background rates, 
would certainly be one of the major tasks for the 2012 runs. In such a case an option of changing 
of the beam energy, proposed in this note,  appears to be clearly superior with 
respect to continuing taking data at the current beam-energy. 

The two scenarios discussed in this note, even if having comparable signal/background resolving power, 
are all but equivalent, as far as the safety of the machine operation is concerned. 
From the machine operation point of view reducing the beam energy to 2.5 TeV represents 
a viable technical solution. The only price to pay would be to accept a slightly diminished sensitivity 
of such runs to the discovery physics at the highest mass scales. This price depends 
upon the evolution of the machine luminosity in the year 2012. If a plateau of the 
instantaneous luminosity is reached by the time of collecting 6  fb$^{-1}$ at 3.5 TeV, the impact of the 
expected increase of the sample of events collected by the end of 2012 at the same energy
both for the searches and for the SM measurements  would be marginal. 
In our view, a change of the beam energy  would be superior with respect to 
continuing running present energy because of several other reasons\footnote{The most notable gain would be to 
control experimentally the contribution of higher twists to the LHC observables -- the domain where 
the theoretical calculations and modeling tools hardly exists.}.  

The option of increasing the LHC beam energy to 4.5 TeV in 2012 is another story. 
We are fully aware that running a  4.5 TeV proton beam before the 2013/2014 shutdown may simply be impossible  because of machine safety arguments\footnote{We evaluated as well perhaps a more realistic  scenario of collecting 
8 fb$^{-1}$ with the 3.5 TeV beams and 7  fb$^{-1}$ at 3.97 TeV. 
We found that the resolution power of the quark- and the gluon-initiated processes, 
for this running scenario is reduced  by a factor of $\sim 2$.  
Such a running scenario, for  which $s_1/s_0 = M_Z^2 / M_W ^2$, 
would fulfill  a double role: in addition to the one discussed in this paper it would be 
crucial for the competitive precision measurements of  the $W$-boson mass and $\alpha _s$\cite{krasny_COPIN}.}.
We were prompted  to include in our paper  the calculations for the increased beam energy 
by the statement of S. Myers at the June 2011 session of the LHCC\cite{Myers}: 
`` Following measurements of the copper stabilizers
resistances during the Christmas stop, we will re-evaluate the
maximum energy for 2012 (Chamonix 2012)".

\section{Conclusions}

It is argued that at the LHC a change of beam energy may provide a useful tool to discriminate between production processes in cases where model uncertainties outweigh the gain from statistical error reduction. The argument is applied to the specific case of the $l \nu l \nu$ excess events observed by ATLAS and CMS at the LHC. 
The  presented case study is a concrete example of a complementary approach 
to searches at the LHC which rely on the dedicated measurement procedures rather
then on the specific theoretical  models. Such an approach could be of use 
in  an  advanced  phase of the LHC experimental programme 
when  the ``Promised Land'' of discoveries, precisely chartered  by 
the present theory paradigms, turns out to be a mirage.   


%
%

\end{document}